Anomalous Expansion of Coronal Mass Ejections during Solar Cycle 24 and its Space Weather Implications

Nat Gopalswamy, NASA Goddard Space Flight Center, Greenbelt, MD, USA

Sachiko Akiyama, Seiji Yashiro, Hong Xie, and Pertti Mäkelä, The Catholic University of America, Washington, DC, USA

Grzegorz Michalek, Astronomical Observatory of the Jagiellonian University, Krakow, Poland

Corresponding author: N. Gopalswamy, Code 671, Solar Physics Laboratory, NASA Goddard Space Flight Center, Greenbelt, MD 20771, USA (nat.gopalswamy@nasa.gov)

Key Points
- Cycle-24 CMEs expand anomalously due to the reduced ambient pressure
- The expansion results in weak ICME magnetic field, hence weak magnetic storms
- Weak ambient magnetic field reduces efficiency of SEP acceleration by shocks

**Abstract:** The familiar correlation between the speed and angular width of coronal mass ejections (CMEs) is also found in solar cycle 24, but the regression line has a larger slope: for a given CME speed, cycle 24 CMEs are significantly wider than those in cycle 23. The slope change indicates a significant change in the physical state of the heliosphere, due to the weak solar activity. The total pressure in the heliosphere (magnetic + plasma) is reduced by ~40%, which leads to the anomalous expansion of CMEs explaining the increased slope. The excess CME expansion contributes to the diminished effectiveness of CMEs in producing magnetic storms during cycle 24, both because the magnetic content of the CMEs is diluted and also because of the weaker ambient fields. The reduced magnetic field in the heliosphere may contribute to the lack of solar energetic particles accelerated to very high energies during this cycle.

## 1. Introduction

Coronal mass ejections (CMEs) have been established as the primary source of major geomagnetic storms and large solar energetic particle (SEP) events [see e.g., Gosling, 1993; Reames, 1999; 2013; Gopalswamy *et al.* 2004; Mewaldt 2006; Zhang *et al.* 2007]. Although the number of large SEP events during cycle 24 is similar to that of cycle 23, the highest energy SEP events and major geomagnetic storms have become rare [Gopalswamy, 2012]. The situation remains the same as of this writing: Table 1 shows the updated properties of major space weather events compared between cycles 23 and 24. There are three striking observations one can make from Table 1: (i) The number of major geomagnetic storms (Dst ≤ -100 nT) is by a factor >3 lower during cycle 24 compared to the corresponding epoch in cycle 23 (May 1996 – January 2014), (ii) There is a dearth of ground level enhancement (GLE) in cycle-24 SEP events, although the number of large SEP events (proton intensity ≥10 particle flux units in the >10 MeV channel) is not too different from that in cycle 23, (iii) The speeds and halo fractions of the CMEs in cycle 24 causing major space weather events (large storms and SEP events) are generally higher than are the case for corresponding CMEs during cycle 23 (halo CMEs appear to fully surround the coronagraph's occulting disk in projection [Howard *et al.* 1982]). These indicate that a faster/wider CME is required to produce a significant space weather event in cycle 24 than in cycle 23 and that the reduced numbers of storms and GLEs in cycle 23 may not simply be due to the approximate factor-of-two difference in the size of the cycles as indicated in Table 1. Since CMEs are the source of these major events, we examined the CME properties during solar cycles 23 and 24 to further investigate the cause(s) for the mild space weather [e.g., Richardson, 2013] during solar cycle 24. In particular, we consider the relationship between the speed and width of CMEs, which are the basic attributes that organize space weather events [see e.g., Gopalswamy *et al.* 2010a]. The motivation for this study came from the result that all SEP-associated CMEs are full halos during cycle 24, compared to about 60-70% in cycle 23 [Gopalswamy, 2012].

**2. Data Selection**

The Large Angle and Spectrometric Coronagraph [LASCO, Brueckner *et al.* 1995], on board the Solar and Heliospheric Observatory (SOHO) mission, has been observing CMEs from the end of cycle 22 to date. This uniform and extended CME data base has become critical in understanding the long-term eruptive behavior of the Sun [Gopalswamy *et al.* 2009a]. We make use of the CME measurements compiled and made available online [http://cdaw.gsfc.nasa.gov/CME_list, see Gopalswamy *et al.* 2009a]. Figure 1 shows the CME daily rate averaged over Carrington Rotation (CR) period for the rise phase of solar cycles 23 and 24. The peak rate during the maximum is similar for the two cycles and the average rate over the study period is the same in the two cycles (1.95 per day, see Table 1). As noted in Table 1, however, the sunspot number (SSN) is significantly smaller during cycle 24.

We investigate the speed and angular width of CMEs, selected by the following criteria: (1) the CMEs must be associated with soft X-ray flare size ≥ C3.0, and (2) the CMEs must originate within $30^o$ of the limb (as determined from the flare location). The first criterion eliminates ambiguities in CME identification for weak flares and the second one minimizes projection effects in speed and angular width measurements. We started with all ≥ C3.0 flares, reported by NOAA's Space Weather Prediction Center with known source locations on the Sun and retained only the ones that occurred within $30^o$ of the limb. We then compiled the associated CMEs from the SOHO/LASCO CME catalog. For CMEs not yet listed in the catalog, we measured their speed and width from the LASCO images. We also independently verified the CME locations using white light and EUV images and movies obtained by SOHO and the Solar Terrestrial Relations Observatory (STEREO) to confirm the physical connection between the flares and CMEs.

The study interval is specified by the length of observations available in cycle 24, from December 1, 2008 to January 31, 2014 (62 months). The corresponding 62-month epoch in cycle 23 is from May 10, 1996 to July 10, 2001. When we refer to cycle 23, it corresponds to the first 62 months of the cycle unless stated otherwise. During these intervals, 148 (cycle 24) and 230 (cycle 23) CMEs were identified, which we use for this study. Even though the CME rate seems to be similar in the two cycles, the overall level of energetic events is smaller during cycle 24. This is clear from the sample sizes over the same time interval in the two cycles (see Table 1): 36% fewer events in cycle 24 (148 vs. 230). The CME speeds (V) ranged from ~100 km/s to >2500 km/s, with similar mean values: 645 km/s (cycle 23) and 685 km/s (cycle 24). The CME width (W) ranged from ~$10^o$ to >$120^o$. The data include 9 full halos (W=$360^o$) in cycle 23 compared to 20 (or 14%) in cycle 24. The number of partial halo CMEs ($120^o$ ≤W<$360^o$) were similar: 46 (or 20%) in cycle 23 compared to 41 (or 28%) in cycle 24. Regular CMEs (W<$120^o$) were more numerous in cycle 23 (175 or 76%) compared to those in cycle 24 (87 or 58%).

The average and median widths of cycle 24 CMEs deviate from the corresponding ones in cycle 23. For regular CMEs (W<120°), the mean and median widths were 58°.5 and 55° (cycle 23) compared to 61°.6 and 60° (cycle 24). When full halos were excluded (W < 360°) the mean and median values were 82°.5 and 69° (cycle 23) compared to 98.1° and 84° (cycle 24). When all CMEs were included, the mean and median values became 93°.4 and 70°.5 (cycle 23) compared to 133°.5 and 98° (cycle 24). Throughout this paper, "halo" refers to full halo CMEs (W=360°) unless specified otherwise. The significant difference in widths is further quantified in the next section.

**3. Speed-Width Relationship of CMEs**
Employing selection criteria similar to those used here, Gopalswamy *et al.* (2009b) showed that V and W are correlated with a correlation coefficient (r) of 0.69 for cycle-23 CMEs. The V - W scatter plots in Fig. 2 confirm this correlation, but the slopes of the regression lines are remarkably different: W = 0.11V + 24.3 for cycle 23 (r = 0.63) compared to W = 0.16V + 24.6 for cycle 24 (r = 0.72). Clearly, the cycle-24 regression line is steeper by ~46%. i.e., for a given CME speed, the cycle-24 CMEs are wider. For V = 1000 km/s, we see that the cycle-24 CMEs are wider by ~38%. Based on regression analyses, the null hypothesis that the two slopes are the same can be rejected. The probability p that the slopes are the same by chance is <0.0009.

The Kolmogorov-Smirnov test (KS-test) showed that the speed distributions of cycles 23 and 24 are not significantly different (p=0.58). Since CME speeds have a log-normal distribution [Yurchyshyn *et al.* 2005], we performed the Student's t-test on the logarithms of speeds, which also showed that there is no significant difference between the means of the two distributions: the mean speed is $10^{2.71\pm0.30}$ km/s and $10^{2.73\pm0.31}$ km/s, respectively for cycle 23 and 24 (p = 0.45). However, the KS test showed that the width distributions are significantly different (p=0.001) between the two cycles. The Student's t-test on the logarithm of widths also confirmed that the width distributions are significantly different (p=0.0002). The means of the width distributions are $10^{1.84\pm0.35}$ degrees and $10^{1.98\pm0.37}$ degrees, respectively for cycle 23 and 24. The 95% confidence intervals for the means are $10^{1.79}$ to $10^{1.89}$ degrees for cycle 23 and $10^{1.92}$ to $10^{2.04}$ degrees for cycle 24. When full halos are excluded, the width distributions are still significantly different (p=0.022). When only regular CMEs are considered (W<120°), the distributions have no significant difference (p=0.43). Thus the difference in the width distributions is consistent with the different speed-width slopes for the two cycles. The higher halo fraction in the cycle-24 population (20/148 or 14%) compared to the 4% (9/230) in cycle 23 (see Table 1) is also consistent with the slope difference.

**4. Why Do We Have Wider CMEs in Cycle 24?**
It is well known that the CME flux ropes expand as they propagate into the heliosphere due to the fact that the total external pressure (magnetic + plasma) declines with the distance from the Sun [see e.g. Shimazu and Vandas, 2002 and references therein]. It appears that the reduced pressure in the coronagraph field of view between cycles 23 and 24 may be responsible for the increase in the CME width. We do not have reliable measurements of the coronal magnetic field in the LASCO FOV. However, we have good measurements of magnetic field, density, and temperature at 1 AU, which we can use to compute the total pressure. We can then extrapolate the values obtained at 1 AU to the coronagraph FOV (~20 Rs) near the Sun.

Figure 3 shows the total pressure ($P_t$) at 1 AU and at the Sun computed from the measured and extrapolated quantities (magnetic field strength B, proton density $N_p$, and proton temperature $T_p$), respectively. The measured quantities are monthly averages from January1996- January 2014 obtained from NASA's OMNIweb (http://omniweb.gsfc.nasa.gov/). The magnetic pressure $B^2/8\pi$ is added to the plasma pressure $N_i k_B T_i + N_e k_B T_e$ to get $P_t$. The subscripts i and e correspond to ions and electrons, respectively. For simplicity, we assume that $T_e = 1.3 \times 10^5$ K and the temperature ($T_\alpha$) of alpha particles is four times $T_p$. We also take $N_\alpha = 0.04\ N_p$ and Ne = 1.04 $N_p = N_p + N_\alpha$ [see e.g., Jian *et al.* 2006]. We see that the 1-AU total pressure during cycle 24 maximum is well below the peak value during cycle 23 (by ~40%). $P_t$ averaged over the maximum phase (2000-2002 in cycle 23 and 2011-2013 in cycle 24) dropped by 24%. The lowest $P_t$ occurred during the cycle 23/24 minimum. The increase during the rise phase of cycle 24 is at the level of cycle 22/23 minimum. The extrapolated $P_t$ near the Sun behaves in a similar fashion, suggesting that CMEs of cycle 24 are ejected into an ambient medium of much lower pressure. The reduced pressure allows the CMEs to expand more, resulting in the higher width for a given CME speed. The higher halo fraction in various CME populations noted in Table 1 can be explained by this anomalous expansion because halo CMEs expand more rapidly early on and appear to surround the occulting disk [Gopalswamy *et al.* 2010b]. We also note that the $P_t$ and B variations are somewhat similar and have implications for space weather, as discussed in the next section.

**5. Implications for the Mild Space Weather**
The remarkable change in the speed-width relationship between cycles 23 and 24 has important implications for the milder space weather in cycle 24, such as weaker and less frequent geomagnetic storms and lower energy SEP events. The minimum value of the Dst index is known to be related to the speed (V in km/s) and the southward component of the magnetic field ($B_z$ in nT) of the interplanetary CMEs (ICMEs): Dst = -0.01V$B_z$ -32 nT [Wu and Lepping, 2002; Gopalswamy, 2010]. Thus for a given ICME speed, the Dst value depends on the magnitude of $B_z$. During cycle 24, there were only 11 large storms (Dst ≤ -100 nT) compared to 40 in cycle 23 over the same epoch. In cycle 24, the storm-causing $B_z$ period was found in CIRs (1), ICMEs (5), and shock sheaths (5). The corresponding numbers in cycle 23 were 4 (CIRs), 22 (ICMEs) and 14 (sheaths). Thus there were nearly four times more ICME-related storms in cycle 23. Figure 4 shows that the average $B_z$ in cycle-24 ICMEs was only 17.3 nT compared to 20.1 nT for cycle 23. Note that no storms had associated fields > 20 nT in Cycle 24 (vs. 8 in cycle 23), although extreme events such as the backside event of 2012 July 23 with a Bz of ~50 nT can still occur [Baker et al. 2013]. There were only 5 major storms in cycle 24 caused by ICMEs (the other five were due to sheaths), so a statistical analysis is not valid. The reduced number of storms can be attributed to the reduced number of energetic CMEs and the lack of large $B_z$ values. Since disk-center fast (V ≥ 750 km/s) and wide CMEs (W ≥ 60°) are generally responsible for large geomagnetic storms, we counted the number of CMEs originating within a central meridian distance of 30° in the two cycles. There is a 47% reduction of such energetic events from cycle 23 (38 CMEs) to cycle 24 (20 CMEs), which alone cannot explain the 72% (36 to 10) reduction in the number of major storms. The anomalous CME expansion is expected to reduce the ICME magnetic field strength and, hence, cause weaker storms. The reduced ambient field strength is also likely to result in weaker compressed sheath fields, and hence, less frequent/weaker sheath storms. It is well known that north-south (NS) ICMEs are more abundant during even-numbered cycles such as 24. This may not be the reason for the diminished geoeffectiveness of the ICMEs because NS types are as effective in producing geomagnetic storms as the south-north ones [Gopalswamy, 2008 and references therein]. In the five ICME storms reported in this paper, only one was due to an NS type ICME. The remaining four were due to high-inclination ICMEs with a south-pointing flux rope axis.

The number of large SEP events in cycle 24 (31) is similar to that in cycle 23 (37). However, only two GLE events have been observed in cycle 24 compared to 7 in cycle 23 (see Table 1). The paucity of GLE events suggests that particles were not accelerated to very high energies. To firm up this conclusion, we examined GOES high energy proton data up to >700 MeV (http://satdat.ngdc.noaa.gov/sem/goes/data/new_avg/). For both cycles 23 and 24, only the GLE events had a discernible signal in the >700 MeV channel. We then examined the proton intensity in the 510-700 MeV channel for all large non-GLE SEP events in cycles 23 and 24. We found three in cycle 24 (2012 January 27, 2012 March 07, and 2013 May 22) and five in cycle 23 (1997 November 4, 1998 November 14, 2000 November 24, 2001 April 2, and 2001 April 12). Thus, there were 12 >500 MeV SEP events in cycle 23, compared to 5 in cycle 24. We can conclude that particles were accelerated to high energies in a smaller fraction of cycle-24 SEP events (5/31 or 16% vs. 12/37 or 32%). We also counted the fast and wide CMEs from GLE longitudes (W20-W90), yielding 55 and 43 CMEs in cycles 23 and 24, respectively. The reduction from 55 to 43 (or 22%) is again not adequate to explain the 62% drop in >500 MeV SEP events. The lack of high-energy particle events can also be attributed to the unfavorable physical conditions in the ambient medium and lack of latitudinal connectivity to the observer [Gopalswamy *et al*. 2013]. The slight reduction in the Alfven speed in the ambient medium (from ~650 km/s in cycle 23 maximum compared to 540 km/s in cycle 24 – see Fig. 3) makes it easier to form shocks, and hence might explain why the number of cycle-24 NOAA-class (>10 pfu at >10 MeV) SEP events did not drop significantly. However, the lack of GLE events seems to suggest that either the particles are not accelerated to very high energies or the source region is not magnetically connected to the observer. Evidence for both of these possibilities was presented in Gopalswamy *et al*. [2013]. While a detailed analysis of the acceleration mechanism is beyond the scope of this paper, we note that the reduced field strength in the ambient medium is likely to reduce the efficiency of shocks in accelerating particles to high energies. This is related to the key role played by the ion Larmor radius in the shock acceleration process. For quasi-parallel shocks (in which the diffusive shock acceleration dominates) and quasi-perpendicular shocks (in which the shock drift acceleration dominates), the energy gain rate of ions for a given shock speed is proportional to the gyrofrequency and, hence, the ambient the magnetic field: $dE/dt \propto B$ [see e.g., Kirk, 1994; Giacalone, 2013]. Therefore, particles may not gain high energies within the available time when the magnetic field strength is low.

# 6. Conclusions

The primary finding of this work is that CMEs of cycle 24 expand anomalously compared to those in cycle 23 as illustrated by the change in slope of the CME speed-width relationship.  For a given CME speed, the cycle-24 CMEs are significantly wider. This is also supported by a larger fraction of halos among cycle-24 CMEs.  The anomalous expansion of CMEs can be attributed to the significant reduction in the observed total pressure (magnetic + plasma) in the ambient medium into which the CMEs are ejected.  We propose that the anomalous CME expansion and diminished ambient solar wind fields in cycle 24 led to the observed reduction in the frequency of large geomagnetic storms via CME field dilution and weaker compressed sheath fields. The reduced Alfven speed in the corona makes it easier to form shocks, which can explain why the number of large SEP events did not drop significantly in cycle 24. Finally, we suggest that the reduced ambient magnetic field strength might have reduced the efficiency of particle acceleration by shocks in cycle 24 and hence might have contributed to the paucity of GLE events.

**Acknowledgments:** SOHO is a project of international cooperation between ESA and NASA. This research was supported by NASA LWS TR&T program. We thank E. W. Cliver and an anonymous referee for their constructive criticism, which improved the presentation of the paper.

**Figure Captions**

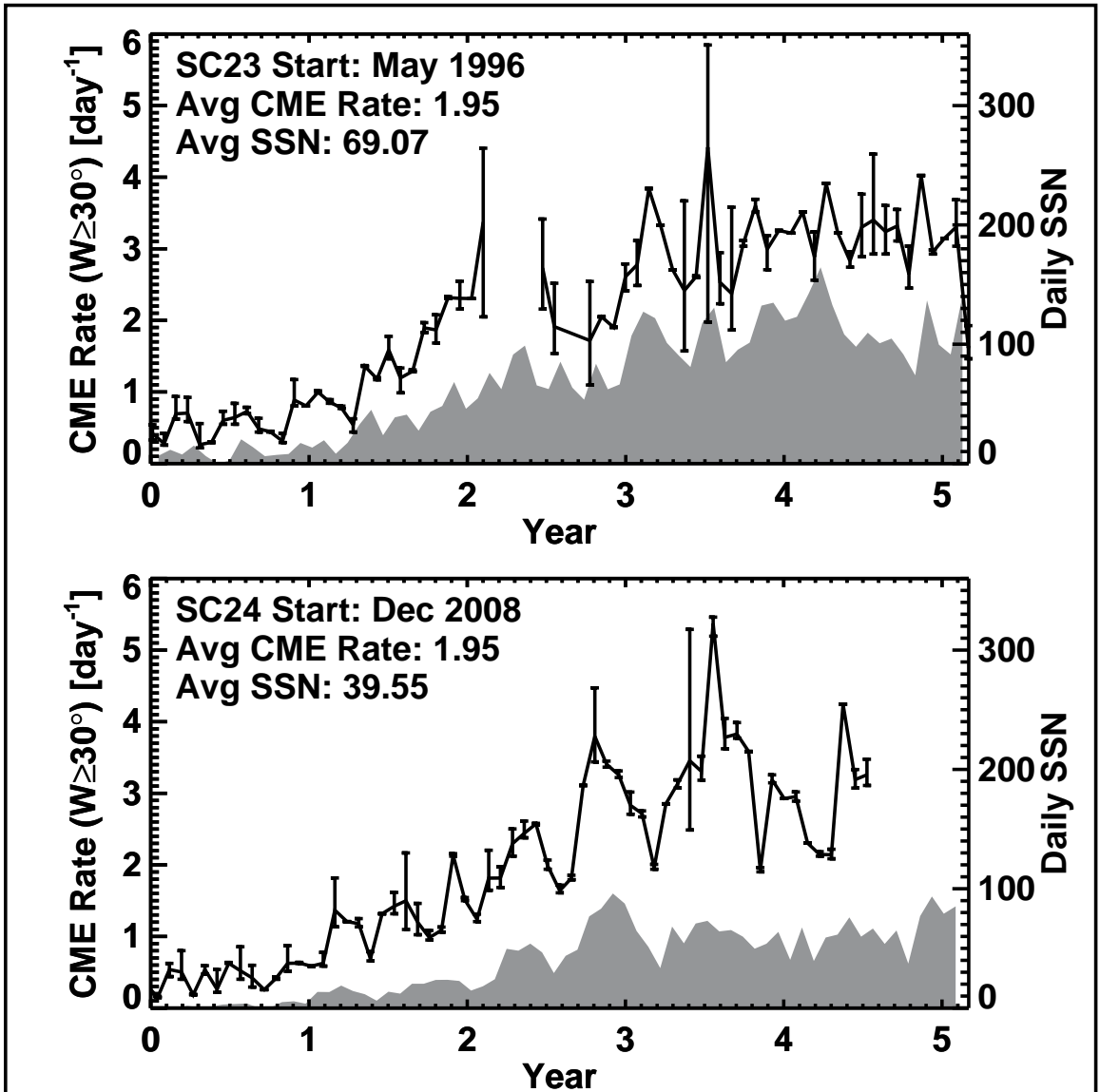

**Figure 1.** CME daily rate averaged over Carrington Rotation (CR) period (27.24 days). The error bars are derived from the cumulative down time in a given CR, counting data gaps (due to roll maneuver and LASCO electronics box anomalies) longer than 3 h. The upper limit was obtained assuming that during the downtime, CMEs occurred at the maximum daily rate of the CR. The lower limit was obtained assuming that no new CMEs occurred during the downtime. The plot in gray is the daily international sunspot number (SSN) obtained from Solar Influences Data Center (http://sidc.oma.be/sunspot-data/). For definitiveness, we have included only CMEs with the width (W) $\geq 30°$. This will eliminate the uncertainty due to narrow CMEs from the disk center missed by the coronagraph. Cycles 23 and 24 overlapped during the year 2008. CME measurements are not yet available after July 2013.

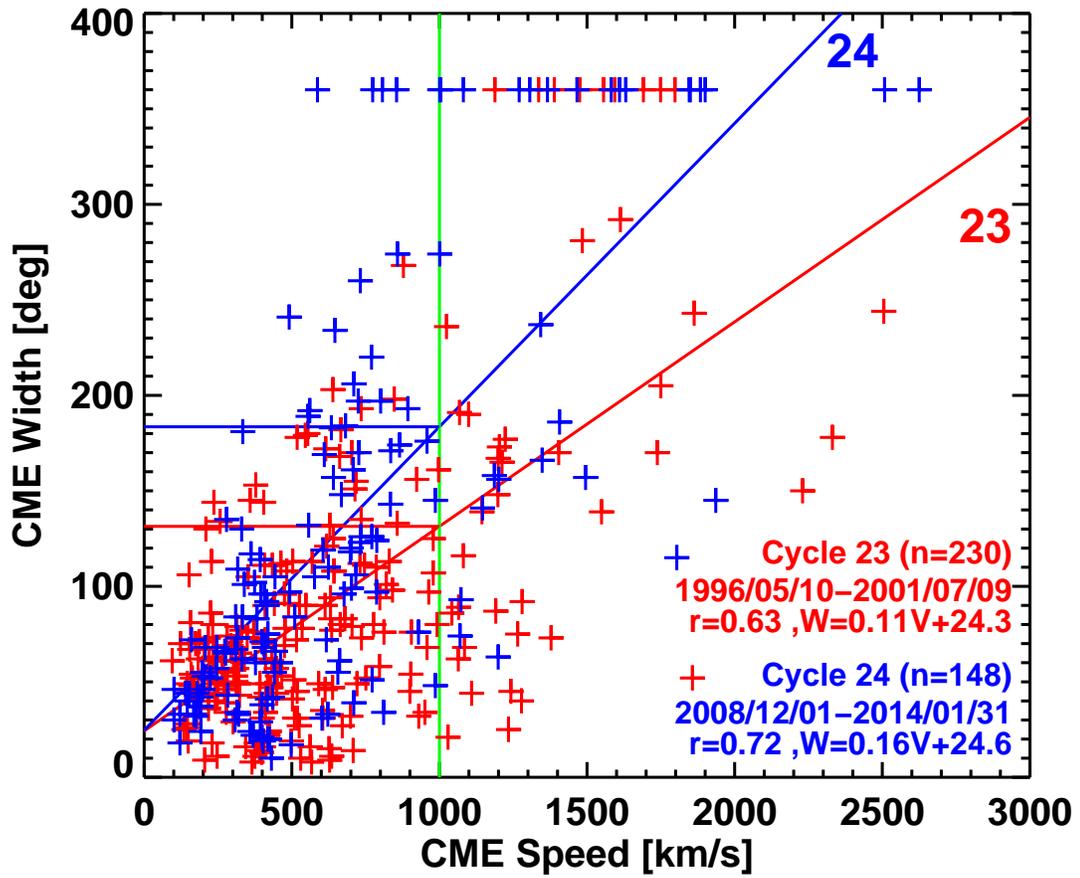

**Figure 2.** Scatterplots between CME speed (V) and angular width (W) for cycles 23 (red) and 24 (blue). The regression lines and the correlation coefficients are indicated on the plot. The width difference at 1000 km/s (green line) is substantial between the two cycles.

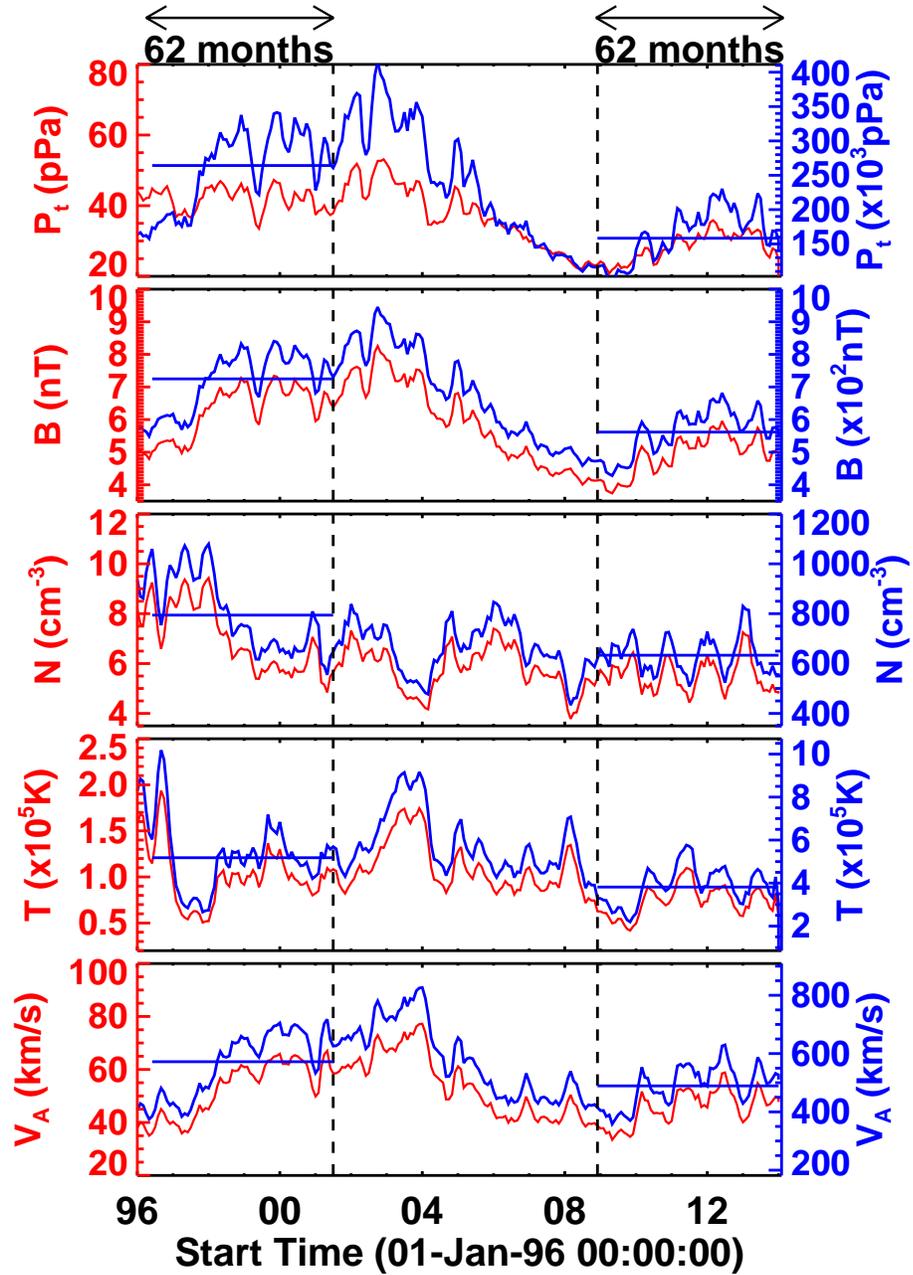

**Figure 3.** Total pressure ($P_t$), magnetic field magnitude (B), proton density (N), proton temperature (T), and the Alfven speed ($V_A$) at 1 AU obtained from OMNI data (red lines with left-side Y-axis). The same quantities extrapolated from 1 AU to the coronagraph FOV (20 Rs) are shown by blue lines (right-side Y-axis). We assumed that B, N, and T vary with the heliocentric distance R as $R^{-2}$, $R^{-2}$, and $R^{-0.7}$, respectively. The blue bars denote the 62-month averages in each panel, showing the decrease of all the parameters in cycle 24.

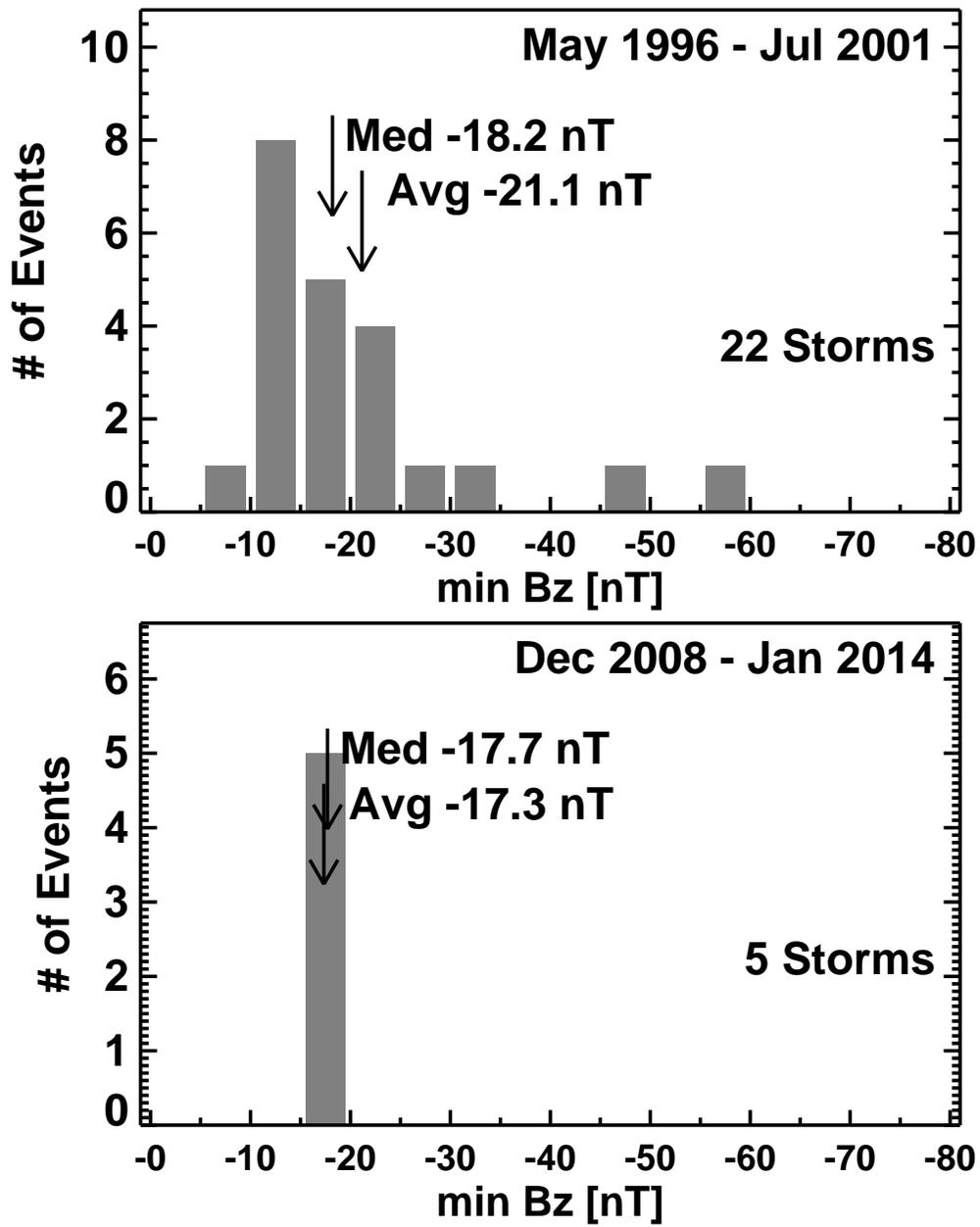

**Figure 4.** Distribution of $B_z$ in major storms of cycle 23 (top) and cycle 24 (bottom).

**Table 1.** Properties of major space weather events during cycles 23 and 24

| Property | Cycle 23[a] | Cycle 24[a] | Ratio[k] |
|---|---|---|---|
| Average Sunspot Number (SSN) | 69.07 | 39.55 | 0.57 |
| Average CME daily rate | 1.95 | 1.95 | 1.0 |
| Number of large magnetic storms[b] | 40 | 11 | 0.28 |
| Number of Sheath storms | 14 | 5 | 0.36 |
| Number of ICME storms | 22 | 5 | 0.23 |
| Number of CME-related storms (Sheath + ICME) | 36 | 10 | 0.28 |
| Average CME speed (magnetic storm) | 722 km/s[c] | 1025 km/s | 1.42 |
| Halo CME fraction (magnetic storm) | 61%[c] | 70% | 1.15 |
| Number of large SEP Events[d] | 37 | 31 | 0.84 |
| Number of SEP events with >500 MeV particles[e] | 12 | 5 | 0.42 |
| Number of GLE Events | 7 | 2[f] | 0.29 |
| Average CME speed (SEP) | 1425 km/s[g] | 1533 km/s | 1.08 |
| Halo CME fraction (SEP) | 65%[g] | 100% | 1.54 |
| Number of CMEs (this work)[h] | 230 | 148 | 0.64 |
| Average CME speed (this work) | 658 km/s | 688 km/s | 1.05 |
| Halo CME fraction (this work) | 4% | 14% | 3.5 |
| Number of fast and wide CMEs (disk center)[i] | 38 | 20 | 0.53 |
| Number of fast and wide CMEs (W20-W90)[j] | 55 | 43 | 0.78 |

[a] May 1996 to July 2001 (cycle 23) and December 2008 to January 2014 (cycle 24)
[b] Dst $\leq$ -100nT for large storms; includes CIR storms (4 in cycle 23 and 1 in cycle 24)
[c] CME measurements available for 30 events; 6 events occurred during SOHO data gap
[d] Proton intensity $\geq$ 10 pfu in the >10 MeV GOES energy channel; 1 pfu = 1 particle per $(cm^2.s.sr)$.
[e] GOES proton data available at http://satdat.ngdc.noaa.gov/sem/goes/data/new_avg/
[f] GLE event on 2014 January 06 was detected only by South-pole Neutron Monitors; however, GOES proton flux shows enhancement in the >700 MeV energy channel.
[g] CME measurements available for 31 events; 6 events occurred during SOHO data gap
[h] CMEs associated with soft X-ray flare size $\geq$C3.0 originating from within 30° of the limb.
[i] CMEs Fast ($\geq$ 750 km/s) and wide ($\geq$60°) associated with soft X-ray flare size $\geq$C3.0 originating within 30° from the central meridian.
[j] CMEs Fast ($\geq$ 900 km/s) and wide ($\geq$60°) originating in the longitude range W20-W90; no restriction on the flare size.
[k] Ratio of cycle 24 to cycle 23 values.